# Is Faith The Enemy Of Science?

Richard MacKenzie

Physique des particules, Université de Montréal

C.P. 6128, Succursale Centreville, Montréal Qc H3C 3J7

richard.mackenzie@umontreal.ca

Abstract: In this article, inspired by Lawrence Krauss' plenary talk at the 2007 CAP congress, the relation between faith and science is examined.

Some readers of this article may have seen Lawrence Krauss' entertaining and thought-provoking plenary talk at the 2007 CAP congress in Saskatoon. They may remember that the first individual to raise a hand during the question period did so in an almost-Horshackian fashion [1], and that he bluntly asserted that he disagreed strongly with the following statement made by Krauss on the challenge of teaching science to the public:

> Faith is not the enemy.
> Ignorance is the enemy.

That brazen individual was me, and I would like to explain why I disagree with this statement – to the point where I would be inclined to go so far as to interchange the words "faith" and "ignorance."

Let us recall that Krauss' talk, entitled "Selling Science to Unwilling Buyers," concerned itself with how to teach science to the general public – the unwilling buyer – by discussing applications of science which interest the buyer, rather than more traditional "bottom-up" approaches to teaching science which might themselves be partly responsible for the preconception that science is boring in the first place. Krauss talked about the current (and abysmal) state of scientific illiteracy in the US, its underlying causes, and how to improve the situation. In giving such a talk to an audience of physicists, Krauss was to a large extent preaching to the converted, and I myself was entirely in agreement with everything he said, except for the above quote, which I found surprisingly discordant with the rest of his talk.

Let me begin with working definitions for the two key terms in the discussion, "science" and "faith."

Science is the study of natural phenomena in order to understand and explain the world around us. This can be with the ultimate goal of serving humanity with such technological advancements as the wheel, agriculture and television (presumably reality TV was not foreseen by those who imagined that television would improve the human condition), or it can simply be for the sake of understanding how the world around us works.

Byron Jennings [2] went into far greater detail on the nature of science than I intend to do here, but the fundamental tool of science is the scientific method: making careful observations or performing controlled experiments, developing a theory which explains the results, making predictions from the theory, and putting the theory to the test with further observations. While this is no doubt a simplistic view of how scientists do science, and there are surely many notable exceptions, it is probably an accurate representation of the way the majority of scientists work.

By "faith" I mean the unsubstantiated belief in something. Now, "unsubstantiated" is somewhat subjective and not a binary concept, but let us not get too distracted by pedantic details. The "something" can be just about anything: that the Montreal Canadiens are going to win the Stanley Cup, that all matter and energy is made up of tiny, vibrating strings, or that there is a benevolent higher power who loves us and is watching over us, but who regularly tests our faith by throwing cataclysmic events at us which leave hundreds of thousands of us dead or in despair. (These are of course minor inconveniences compared to the Sun's evolution towards a red giant that awaits us in a few billion years or so.)

These three examples illustrate the breadth of the term faith (or my definition of it, at least). Belief that the Canadiens are going to win the Stanley Cup is not a statement about reality today, but about a future reality. There may be some measure of justification for holding this belief, and the justification may be compelling or not, so this is perhaps not the greatest example of *unsubstantiated* belief. But ultimately the Canadiens will either win the Stanley Cup or not, leaving the believer either gloating because s/he knew something the rest of us didn't, or quietly forgetting that s/he had ever held that belief in the first place.

Belief in string theory, in contrast, is a statement about an ongoing, for all intents and purposes eternal, reality. Matter either is or is not made of strings; as far as I know, no one believes that this description came into being last week, or that it will no longer be valid next week.

Is belief in string theory unsubstantiated? Again, this is a tricky point. Some have argued for a much stronger statement: that string theory is not only unsubstantiated but untestable – that it will always be unsubstantiated. This point of view is often portrayed as "demoting" string theory to "mere" religion or philosophy. I will argue against both points (although I will fall far short of the opposing view held by a certain Canadian theoretical physicist who, tongue in cheek, drops a pencil to "prove" the validity of string theory).

First, while there is no direct evidence for string theory, there are legitimate scientific reasons for taking it seriously as a description of the world around us. The fact that it is a quantum theory of gravity makes it worth consideration, in spite of a few minor remaining hurdles to overcome such as explaining the observed dimensionality of space-time.

Second, even if sceptics consider the preceding argument a bit thin, string theory is certainly not fundamentally untestable. Eventually string theory will presumably make actual predictions (if this were not the case, then I would agree with its demotion to religion or philosophy), and its validity will either be supported by or contradicted by future observations or experiments. In the former case, string theory will not have been *proven* to be the correct description of reality, but its credibility will take a giant leap upward, leaving many erstwhile sceptics scrambling to understand why string theorists draw what to the untrained eye appear to be bagels on the blackboard and describe them as compactified AdS geometries. In the latter case, string theory must be rejected, or modified in such a way as to bring it into agreement with the "offending" observations or experiments.

The third example of faith is quite different. Like the second, it is a statement about an eternal reality; however (as far as I can see, at least) it is not a belief that will one day be tested (unless, of course, the higher power sends an indisputable sign our way). This being the case, it is perhaps the gold standard in *unsubstantiated* belief. I imagine it is this type of faith – whether you call it religious faith, belief in God, or whatever – that Krauss had in mind in his talk, and that most of us associate with the word "faith" in any case, so I will restrict myself to this definition of faith in what follows.

What, then, is the relation between faith and science? Put simply, in my mind they are diametrically opposed to one another. The cornerstone of faith is blind, unquestioning acceptance; that of science is observation and scrutiny – and the willingness to accept that a cherished belief is wrong if it conflicts with observation. It's hard to imagine two world-views more different than that.

I get a powerful illustration of the contrast between the two when (as happens every few months or so) the Jehovah's Witnesses come to my door to promote their religion. I am perhaps in a minority in that I actually enjoy talking to them; the conversation is always very civil and respectful, although to date they have not succeeded in enlightening me (nor I them). At some point, I get around to telling my visitors that I would be only too happy to believe in God, but that, despite considerable reflection, I can't think of a single reason why I should. To them, all the proof I could possibly need is in the Bible, whose many predictions have all come true with absolute precision. (I was once offered the following example: that the world will be filled with misery and evil.) After they have assured me that they stand by every word in the Bible (since, after all, it *is* the word of God), I question them about the fact that the Universe is known to be billions of years old and not thousands… and the back-pedaling begins: the word of God is a moving target.

The reason I bring this up is that I feel I am merely being a scientist and not an atheist when I tell my proselytizing visitors that I need a reason to believe in God. The origin of the Universe is as much a mystery to me as to anyone, and maybe it was created by a Creator… but that is about as far as my minimalist, scientific philosophy will allow me to go without further evidence. To evoke a well-defined, precise notion of God (a humanized notion, if you will) who created the Universe, life, and man in His own image is a huge, unsubstantiated extrapolation that is grotesquely unjustified, nicely illustrated by Bertrand Russell's celestial teapot analogy [3] and, more recently and in a lighter vein, by the Flying Spaghetti Monster [4]. I owe it to myself as a scientist to try to minimize the number of assumptions I make in my philosophy of life, just as almost any scientist does in formulating an explanation for any observable phenomenon. Creator? Maybe. God who, though omnipotent, waited 15 billion years before having a son, according to one widespread religion? I don't think so.

Recently, I was told by a colleague that he believes in God but that he separates this side of his philosophy from the scientist side. Although our conversation went in a different direction, it occurred to me that he might quantify his philosophy of life by a point in a plane whose axes are "scientificity" and "faith" or "religiosity." If the origin is neutral on both counts, I would guess that he would place himself somewhere in the first quadrant, given that he is rather scientific and obviously at least somewhat religious.

In my opinion, only one dimension is necessary, with science to the right, say, and faith to the left. The more a person is a true scientist, the less willing s/he is to accept notions as true without a reason for accepting them, so the less inclined s/he is towards unsubstantiated belief – faith. So already we see that faith and science are antagonistic.

But to what extent is faith the *enemy* of science? That is a trickier point. Does faith obstruct scientists from doing science? Seemingly not; clearly there is no shortage of religious scientists who do good, respectable science, although the combination strikes me as having at least a degree of internal inconsistency. But I would argue that the overwhelming majority of modern science is far from the arena of potential conflict. It simply doesn't matter how strong the researcher's faith is when s/he is removing atoms from the surface of a solid with a laser or studying magnetic currents in the Sun. Science is so incredibly specialized that a student can get a PhD in particle physics without having a clue what supersymmetry is, to say nothing of worrying about how, and for what purpose (if any), the Universe was created.

Indeed, we can go a step farther and observe that even the arena of potential conflict – studying the origin of the Universe, say – is not in conflict with faith, per se (although it is obviously in conflict with a literal interpretation of the Bible). We are back to the observation that religion – faith – is not falsifiable because any scientific observation can simply be argued to be just what God had wanted. In a recent series of articles [5], Don Page argued (among other things) that a current idea in cosmology, the multiverse, like Darwinian evolution, should not be perceived as a threat to Christian theology, although both are widely perceived as such. Given this "non-falsifiability escape clause," Page's conclusion strikes me as rather self-evident, although it depends to a great extent on how malleable the theology is, or in other words on how narrowly *Christian* theology is defined. I personally have a hard time reconciling evolution and even a single Universe (to say nothing of a multiverse) in which our planet is about as remarkable as a grain of sand on a beach, on the one hand, with the idea that humanity is any more important to God than dinosaurs, say, or amoeba, or life elsewhere in the Universe, on the other. Given that the latter seems to be a cornerstone of Christian theology, I think I would feel quite threatened indeed if I subscribed to a Christian theology.

The bottom line is that direct observation shows that faith does not obstruct scientists from doing science. That said, there are many who portray themselves as scientists who, due to their faith, are doing a brand of science which is an indignity to the word. I have in mind particularly those whose principal goal in science is to advance a faith-based agenda. One must wonder whether these individuals, who probably have a reasonable amount of scientific talent, might not be doing respectable science if their scientificity had not been stronger, or their religiosity weaker.

Does faith obstruct non-scientists from learning science? I would argue that it does, for several reasons.

First, there is the issue of what is taught in schools, how it is taught, and by whom. Well-known are the efforts of fundamentalist Christian groups to prevent the teaching of evolution in U.S. schools, and failing that, to advance intelligent design (a close cousin of creationism) as a competing scientific theory which, as such, deserves equal class time – in spite of overwhelming scientific evidence for one and none for the other.("None for the other" is a gross misstatement, in fact, because it implies there could in principle be evidence for it. But creation is by its very nature not a science because it is not falsifiable by observation.) Battles are regularly fought on this issue (and not just in the U.S., by the way: the subject of the teaching of creationism in schools has been in the news in the U.K. and in Canada recently). Science has usually maintained the upper hand, fortunately, but the war goes on.

Second, organizations such as the Seattle-based Discovery Institute and the John Templeton Foundation (both of which have deep pockets) are engaged in a different, more indirect and therefore more dangerous campaign in the "war" against science: to blur the distinction between religion and science. The Discovery Institute is a Christian-based organization with various offshoots, two of which are the Center for Science and Culture and the Biologic Institute. These are pretty scientific sounding names, so one would assume these institutes are engaged in some sort of scientific thinking or research.  But their main mission is to advance the notion of intelligent design as a serious scientific theory in competition with evolution.  The non-scientific public runs the risk of falling into the carefully-laid trap of associating creationism with science.  Further on down the road (if the Discovery Institute gets its way), school boards, voters, and elected representatives at all levels of government will be unsure about what is and isn't science.  If you think this is fearmongering or an overstatement of the facts, just take a look at the man in the White House at the time of this writing.

The Templeton Foundation's slogan is "Supporting science – investing in the big questions." While it has in the past funded activities of the Discovery Institute, to its credit it has more recently distanced itself from intelligent design and from the Discovery Institute itself. Nonetheless, its main thrust can be described as blurring the line between science and religion. The Foundation's flagship offering is a prize whose value is adjusted so that it beats out the Nobel Prize financially (thereby gaining publicity, and probably credibility in the eyes of the dollar-dazed public), for "progress toward research or discoveries about spiritual realities."  Early winners include Mother Theresa and Billy Graham, but the prizes have been more recently been frequently awarded to scientists, including physicists such as Paul Davies and Freeman Dyson.

The third – and in my mind the most important but the least quantifiable and therefore the most contentious – way in which faith gets in the way of non-scientists learning science goes back to the antagonistic relation between faith and science. The blind acceptance that is the hallmark of faith could, if not confined to the arena of one's theology (indeed, why should it be so confined?), compromise one's ability to think critically, a key ingredient in scientific thinking.  I think that in all aspects of life, science and non-science, we must be guided by critical thinking and common sense. (The latter is a tricky issue, as exemplified in the expression "One man's common sense is another's nonsense." I have in mind a sort of intuition or judgment guided by the sort of analytical, scientific thinking we hope to foster in our students.)  This has always been the case, but perhaps it is true now more than ever, given the quantity of demonstrable nonsense on the internet just a click or two of the mouse away. A healthy amount of scepticism helps us evaluate whether we should believe what we hear or read.  Blindly accepting authority is not a very good guideline because an authority can be found to advance virtually any side of any issue, big (whether or not humanity is responsible for global warming), small (whether one should stretch before exercise or afterwards), and everything in between.

Before I accept an idea as valid, it must be scrutinized, or undergo the "common sense test." Should I believe that distant stars and planets have some bearing on my daily life? After thinking about this question for some time, I cannot conceive of a known mechanism for them to do so, so I am doubtful, to say the least: astrology has failed the common sense test. Failing this test is not a proof that an idea is false, of course, but it does place the "burden of proof" on the idea. All religions I can think of fail spectacularly, because even if they are not in direct conflict with scientific observations (the best-known example of this, perhaps, being the Bible), they place great importance on phenomena which have never been observed under controlled conditions (for example, the various [and mutually incompatible] notions of an afterlife found in many common religions).

If one is to adhere to such a system of belief, a sort of Pandora's box is opened. One might as well believe that certain people have the ability to read fortunes, that water (labelled as homeopathic medicine) can cure virtually any illness, that we will be reincarnated after death in a form determined by our behaviour in this life, and so on and so forth. As scientists, we may chuckle at these beliefs, yet they are quite widespread, and are symptomatic of a non-scientific world-view, or, equivalently, of an impaired ability to use the common sense test. This incapacity, in my mind, is a great impediment to understanding the world around us and is an impediment to learning science.

Krauss gave several examples of scientific illiteracy in the U.S., which would be amusing if they were not true. The most spectacular of these was that only about 50% of respondents in a recent survey realized that the Earth goes around the Sun in about a year. Although I don't know if such a test has been done, it would be interesting to see if there was a correlation between religiosity and the answer given for this question.

What of ignorance, which according to Krauss is the enemy of science? Krauss may well have had a more colloquial interpretation of ignorance as a stubborn refusal to accept new ideas, but ignorance actually means a lack of knowledge. In my mind, if the goal is to educate the public, then in a sense Krauss is correct: ignorance is the enemy. However, I am inclined to think that an open-minded but ignorant person, rather than being the enemy, is exactly where the most progress can be made in educating the public, so rather than viewing ignorance as the enemy, I would view it as a primary target, where resources can most effectively be used. It is much easier to write on the clean slate of an ignorant but open-minded person than to have to first erase preconceived notions that run counter to the criticality needed to develop a scientific understanding of the world around us.

In summary, I have argued that faith and science represent antagonistic world-views: blind acceptance is the cornerstone of faith, and is diametrically opposed to the open-minded scepticism of a scientific world-view. I believe that faith can indeed impede non-scientists from learning science, both indirectly (in terms of the choice of school curricula) and directly (since faith by its very nature and definition runs counter to a scientific way of thinking).

Stevie Wonder said it as well as anybody: "When you believe in things that you don't understand, then you suffer. Superstition ain't the way."